\documentclass[%
 reprint,
 amsmath,amssymb,
 aps,
]{revtex4-1}

\usepackage{graphicx}
\usepackage{dcolumn}
\usepackage{bm}
\usepackage{amsmath}    
\usepackage{graphicx}   
\usepackage{hyperref}   
\usepackage{cleveref}   
\usepackage{siunitx}
\usepackage{physics}
\usepackage{aas_macros}
\usepackage{natbib}

\usepackage[table,svgnames,dvipsnames]{xcolor}
\usepackage[mathlines]{lineno}
\usepackage[normalem]{ulem}

\usepackage{url}         
\usepackage{aas_macros}  

\providecommand{\Eprint}[2][]{\url{#2}}

\usepackage[T1]{fontenc}  
\usepackage[utf8]{inputenc} 
\usepackage[table,svgnames,dvipsnames]{xcolor}
\usepackage[mathlines]{lineno}
\usepackage[normalem]{ulem}

\usepackage{titlesec}

\crefname{figure}{Fig.}{Figs.}     
\Crefname{figure}{Fig.}{Figs.}     

\begin{document}

\preprint{APS/123-QED}

\title{Consistent Initial Conditions for Early Modified Gravity in Effective Field Theory}

\author{Jiaming Pan$^{1}$}
\email{jiamingp@umich.edu}

\author{Meng-Xiang Lin$^{2}$}
\email{mxlin@sas.upenn.edu}

\author{Gen Ye$^{3}$}

\author{Marco Raveri$^{4}$}
 
\author{Alessandra Silvestri$^{3}$}

\affiliation{$^{1}$Department of Physics and Leinweber Center for Theoretical Physics, University of Michigan, 450 Church St, Ann Arbor, MI 48109, USA}

\affiliation{$^{2}$Center for Particle Cosmology, Department of Physics and Astronomy, University of Pennsylvania, Philadelphia, Pennsylvania 19104, USA}

\affiliation{$^{3}$Institute Lorentz, Leiden University, PO Box 9506, Leiden 2300 RA, The Netherlands}

\affiliation{$^{4}$Department of Physics and INFN, University of Genova, Via Dodecaneso 33, 16146, Italy}


\begin{abstract}

Precise initial conditions (ICs) are crucial for accurate computation in cosmological perturbation theory. We derive the consistent ICs for Horndeski theory in the Effective Field Theory (EFT) approach, assuming constant EFT functions at early times. We implement the ICs into the public Boltzmann code \texttt{EFTCAMB}, and demonstrate that the expected early-time behavior of perturbations and Weyl potential can be obtained with theory-consistent MG ICs. We identify significant deviations when comparing Cosmic Microwave Background angular power spectra in MG models obtained with consistent MG ICs versus inconsistent GR ICs. Our findings underline the importance of using accurate, theory-consistent MG ICs to ensure robust cosmological constraints on early MG models.

\end{abstract}

\maketitle


\section{\label{sec:intro}Introduction}

The discovery of the late-time cosmic acceleration~\cite{Riess1998,Perlmutter1999} fundamentally transformed our understanding of the Universe, leading to the establishment of dark energy (DE) as a key component of the cosmological model. The standard model of cosmology, $\Lambda$CDM, adopts the cosmological constant $\Lambda$ as the simplest form of DE. Over the years, numerous alternative models have been proposed, ranging from dynamical dark energy to modified gravity (MG) \cite{Copeland:2006wr, Silvestri:2009hh, Clifton:2011jh}. Nevertheless, subsequent observational data have consistently supported the $\Lambda$CDM framework. 
Recently, the DESI collaboration \cite{DESI2025,DESI:2024mwx} inaugurated the era of Stage IV surveys, offering tantalizing hints of possible dynamics in DE. Particularly intriguing from a theoretical standpoint is the indication of a crossing of the so-called phantom divide—namely, a scenario in which dark energy, initially characterized by an equation-of-state parameter $w < -1$, transitions into the $w > -1$ regime at low redshift.

This spurred an intense investigation of DE and MG candidates that could explain the observations, see e.g.  \cite{Ye:2024ywg,Pan2025,Khoury:2025txd,DESI2025Lodha,Addazi:2025qra,Wolf:2025jed,Ye:2025ulq,Chaussidon:2025npr,Nakagawa:2025ejs,Ye:2024zpk,Silva:2025hxw,Li:2025cxn,Yang:2025mws,Liu:2025mub,Ishak:2024jhs}. 
Most single-field DE/MG models can be studied within the unifying framework of Effective Field Theory (EFT)~\cite{Bloomfield:2012ff,Gubitosi:2012hu}. Cosmological inference in these models requires numerically solving the modified Einstein–Boltzmann equations to produce precise predictions for observables, including cosmic microwave background (CMB), galaxy-clustering power spectra, and weak-lensing.

To date, two public codes implement the EFT formalism in Einstein-Boltzmann solvers.  
The first, \texttt{EFTCAMB}~\cite{Hu2014,Raveri2014}, implements EFT formalism into the public \texttt{CAMB}~\cite{Lewis:1999bs,HowlettCAMB:2012} Boltzmann code; 
the second, \texttt{hi\_class}/\texttt{mochiclass}~\cite{Bellini:2019syt,Zumalacarregui:2016pph,Cataneo:2024uox}, works within the \texttt{CLASS} code. 
We work with the former and focus on the treatment of ICs.

Typically, Einstein–Boltzmann solvers initialize perturbations deep in the radiation era  with initial conditions derived under General Relativity (hereafter we will call these GR ICs)~\cite{Ma:1995ey,Bucher:1999re}, assuming MG effects are negligible at early times. DE/MG models that address cosmic acceleration typically fulfill this assumption and do not seem to require the additional complexity of solving for the specific ICs. However, in early MG scenarios where MG/DE effects are significant even during the radiation era, the GR ICs cease to be a good approximation and can cause differences in cosmological observables, as shown, e.g., in \cite{Lin:2018nxe}. Early MG/DE models are gaining more attention as potential candidates for addressing the Hubble tension, see e.g.~\cite{Braglia:2020auw,Poulin:2018dzj,Poulin:2018cxd,Lin2019,Ye:2020btb,Smith2020,Niedermann:2020dwg,Ye:2024zpk}. More generally, with the advent of precision cosmology, it is desirable to set theoretically consistent ICs.

Motivated by these considerations, in this work we derive the consistent ICs for the Horndeski models within the EFT framework, extending previous results~\cite{Liu:2017oey,Lin:2018nxe} to more general parameter spaces. We focus on the ICs for the adiabatic modes, assuming constant EFT functions, and analyze their impact on cosmological observables. Our results show that using consistent ICs can significantly alter early-time cosmological observables including the CMB angular power spectrum, highlighting their importance for parameter inference in early MG scenarios.

The paper is structured as follows. In \cref{sec:method}, we outline the EFT formalism and detail our methodology to derive the ICs. We then present our results, including the analytical solution of the modified ICs, their validity, and the cosmological implications, in \cref{sec:results}. Finally, we discuss our findings and conclude in the last section. Throughout we use the metric signature $(-,+,+,+)$.

\section{\label{sec:method}Methodology}
\subsection{Effective Field Theory of Dark Energy}

 We write the EFT action of DE in the unitary gauge, where spatial hypersurfaces correspond to uniform field hypersurfaces, following the convention used in \texttt{EFTCAMB}~\cite{Hu:2014oga}, i.e.:
\begin{equation}
  \begin{aligned}
    S_{\text{DE}} &= \int d^4 x \sqrt{-g} \Biggl\{ M_{\text{P}}^2 \Bigl[1+\Omega(t)\Bigr] \frac{R}{2} - \Lambda(t) - c(t) g^{00} \\
    &\quad +\frac{M_{2}^{4}(t)}{2}\bigl(\delta g^{00}\bigr)^2 - \bar{M}_{1}^3(t) \frac{1}{2} \delta g^{00} \, \delta K - \bar{M}_{2}^2(t) \frac{1}{2} \bigl(\delta K\bigr)^2  \\
    &\quad - \bar{M}_3^2(t) \frac{1}{2} \delta K^\mu{}_\nu\,\delta K^\nu{}_\mu + \hat{M}^2(t) \frac{1}{2} \delta g^{00}\, \delta R^{(3)}\\[0.2cm]
    &\quad + m_2(t) \, \partial_i g^{00} \, \partial^i g^{00} \Biggr\} + S_{\text{m}}(g_{\mu\nu}, \Psi_{\text{m}}),
  \end{aligned}
  \label{eq:EFTDEaction}
\end{equation}
where \(M_{\text{P}}\) is the Planck mass,  \(R\) is the Ricci scalar, and \(\delta R^{(3)}\) represents the perturbation of the Ricci scalar of spatial hyersurfaces. Here, \(\delta g^{00}\) is defined through \(g^{00}\equiv-1+\delta g^{00}\), and $\delta K^\mu{}_\nu$ denotes the perturbation of the extrinsic curvature of spatial hypersurfaces and its trace \(\delta K\). The matter action \(S_{\text{m}}(g_{\mu\nu}, \Psi_{\text{m}})\) includes all fields except dark energy.

The \emph{EFT functions} \{$\Omega(t)$, $\Lambda(t)$, $c(t)$, $M_2(t)$, $\bar{M_{1}}(t)$, $\bar{M_{2}}(t)$, $\bar{M_{3}}(t)$, $\hat{M}(t)$, $m_2(t)$\} are free functions of time that multiply all operators that are compatible with invariance under spatial diffeomorphisms of the action.  
Following  \cite{Hu2014,Raveri2014}, we introduce the following dimensionless form of the EFT functions:
\begin{equation}
    \begin{aligned}
          \gamma_1 &= \frac{M_{2}(t)^4}{m^2_0 H^2_0}, &
    \gamma_2 &= \frac{\bar{M}_{1}(t)^{3}}{m^2_0 H_0}, &
    \gamma_3 &= \frac{\bar{M}_{2}(t)^{2}}{m^2_0}, \\
    \gamma_4 &= \frac{\bar{M}_{3}(t)^2}{m^2_0}, &
    \gamma_5 &= \frac{\hat{M}(t)^{2}}{m^2_0}, &
    \gamma_6 &= \frac{m_{2}(t)^{2}}{m^2_0}.
    \end{aligned}
    \label{eq:gamma}
\end{equation}
We will focus on the broad class of Horndeski gravity~\cite{Horndeski:1974wa,Horndeski:2024sjk}, therefore 
demanding the following constraints to avoid higher-order spatial derivatives \cite{2013JCAP...08..025G,2015JCAP...02..018G}: \begin{equation}
    \begin{aligned}
    m_2 &= 0, \\
    \hat{M}^2 &= \frac{\bar{M_2}^2}{2} = -\frac{\bar{M_3}^2}{2}.
    \end{aligned}
    \label{eq:hornpri}
\end{equation} These are equivalent to $2\gamma_5 = \gamma_3 = -\gamma_4$ and $\gamma_6 = 0$, which is the parameter subspace corresponding to Horndeski models \cite{Horndeski:1974wa,Horndeski:2024sjk}.

Additionally, two EFT functions out of $\{\Omega, \Lambda,c\}$ can be eliminated using the Friedmann equations, after having chosen a background cosmology. We will use this to remove $\Lambda$ and $c$, (i.e. deriving them as functions of $\Omega$ and $H(z)$) and work with four free EFT functions, namely $\{\Omega, \gamma_{1,2,3}\}$ . When deriving the ICs, we will take the limit of constant EFT functions, setting
\begin{equation}
    \Omega(a) = \Omega_0, \quad \gamma_i(a) = \gamma_{i,0}\,.
\end{equation}

\subsection{Equation of Motion for Linear Perturbations}

In this section, we present the Einstein equations and Boltzmann equations governing metric and density perturbations on super-horizon scales. These equations are used to determine the ICs for Horndeski models within the EFT framework.

In the action of EFT formalism Eq.~(\ref{eq:EFTDEaction}), the extra scalar degree of freedom of DE/MG models is hidden in the metric, through the appropriate choice of slicing (i.e. uniform field spatial hypersurfaces). To work with an Einstein-Boltzmann solver, one needs to undo this choice and transform to either the  Newtonian or synchronous gauge. In other words,  it is necessary to leave the unitary  gauge by making the Goldstone field $\pi(\mathbf{x})$ explicit via the St\"uckelberg trick
\begin{equation}
  \tau \;\longrightarrow\; \tau + \pi(\mathbf{x})\,,
\end{equation}
where $\tau$ is the conformal time.
ICs are set by taking the early times and super-horizon limit of the set of Einstein-Boltzmann equations that regulate the perturbations of the metric, matter, and the dark energy field. 

In the following, we work in the synchronous gauge
\begin{equation}
    ds^2 = a^2(\tau)\left[ -d\tau^2 + (\delta_{ij}+h_{ij})dx^idx^j\right]\,,
\end{equation}
where only the spatial part of the metric is perturbed, and the scalar components $h$ and $\eta$ are defined as, in Fourier space:
\begin{equation}
    h_{ij}(\tau,\vec{x})=\int\,d^3k\,e^{i\vec{k}\cdot\vec{x}}\left[\hat{k}_i\hat{k}_j\,h(\tau,\vec{k})+(\hat{k}_i\hat{k}_j-\frac{1}{3}\delta_{ij})6\eta(\tau,\vec{k})\right] 
\end{equation}
where $\vec{k}=k\hat{k}$. 
We will denote the density contrast, pressure,  velocity potential, and anisotropic stress of any species $i$ with, respectively,  $\delta_i$, $\delta P_i$, $\theta_i$, and $\sigma_i$. The perturbation in the dark energy field is represented by $\pi$. 

In the early epoch, photons ($\gamma$) and baryons ($b$) are tightly coupled via the Thomson scattering process, and form a single plasma with common velocity
\begin{equation}
\theta_b = \theta_\gamma\,.
\end{equation}
The evolution of the photon fluid is governed by
\begin{align}
\dot{\delta}_\gamma + \frac{4}{3}\theta_\gamma + \frac{2}{3}\dot{h} &= 0, \label{eq:photon1} \\
\dot{\theta}_\gamma - \frac{1}{4} k^2 \delta_\gamma &= 0, \label{eq:photon2}
\end{align}
where an overdot denotes differentiation with respect to conformal time.
For the baryon component, the equations read
\begin{align}
\dot{\delta}_b + \theta_b + \frac{1}{2}\dot{h} &= 0, \label{eq:baryon1} \\
\dot{\theta}_b + \frac{\dot{a}}{a}\theta_b &= 0, \label{eq:baryon2}
\end{align}
with $a$ being the scale factor.

Massless neutrinos evolve according to
\begin{align}
\dot{\delta}_\nu + \frac{4}{3}\theta_\nu + \frac{2}{3}\dot{h} &= 0, \label{eq:neutrino1} \\
\dot{\theta}_\nu - \frac{1}{4} k^2\left(\delta_\nu - 4\sigma_\nu\right) &= 0, \label{eq:neutrino2} \\
\dot{\sigma}_\nu - \frac{2}{15}\left(2\theta_\nu + \dot{h} + 6\dot{\eta}\right) &= 0\,. \label{eq:neutrino3}
\end{align}
.

Cold dark matter (CDM), treated as a pressureless perfect fluid that interacts solely through gravity, is described in the synchronous gauge (with the CDM comoving frame) by
\begin{align}
\dot{\delta}_c &= -\frac{1}{2}\dot{h}, \label{eq:colddark1} \\
\theta_c &= 0, \label{eq:colddark2}\,.
\end{align}

This system of equations is closed with the four linear Einstein equations. The latter are not all independent and can be combined into a reduced set of equations, depending on the context. For the numerical evolution inside Einstein-Boltzmann solvers, one typically reduces them to two first order differential equations for $h$ and $\eta$. We present these in the general form that they assume in EFT, i.e.:
\begin{widetext}
\begin{align}
k\,\dot{\eta} &= \frac{1}{X}\left[
\frac{1}{1+\Omega}\frac{a^2\left(\rho_{m,\nu}+P_{m,\nu}\right)}{m_0^2}\frac{v_{m,\nu}}{2}
+ \frac{k^2}{3H_0}\,F
+ \frac{(U-X)\,k}{6}\,\dot{h}
\right]
\label{eq:hornd1},\\[2ex]
\frac{\dot{h}}{2k} &= \frac{1}{G}\left[
\frac{\mathcal{Q}\,k\,\eta}{\mathcal{H}}
+ \frac{a^2\,\delta\rho_{m,\nu}}{2\,\mathcal{H}\,(1+\Omega)\,k\,m_0^2}
+ \frac{L}{k\,H_0}
\right],
\label{eq:hornd2}
\end{align}
\end{widetext}
where \(X\), \(F\), \(U\), \(G\), \(\mathcal{Q}\), and \(L\) are given in \cite{Hu:2014oga}. When we set the ICs, at early times, we work under the assumption of constant EFT functions; correspondingly, the expressions of \(X\), \(F\), \(U\), \(G\), \(\mathcal{Q}\), and \(L\) simplify significantly, reducing to the expressions in Appendix \ref{app:constEFT}. Here, $m_0^2$ is the Planck mass, \(\rho_{m,\nu}\) and \(P_{m,\nu}\) denote the energy density and pressure of the matter and neutrino components, respectively, and \(v_{m,\nu}\) is the corresponding velocity; its divergence in Fourier space is defined by $\theta_{m,\nu} \;\equiv\; i\,k^{j}\,v_{j}\,$. Here \(\mathcal{H}\) is the conformal Hubble parameter, defined as $\frac{\dot a}{a}$, where the dot indicates $d/d\tau$ here and throughout.

Finally, the dynamical evolution of the St\"uckelberg \(\pi\) field is governed by the following equation
\begin{widetext}
\begin{equation}
 A(\tau,k)\,\ddot{\pi} + B(\tau,k)\,\dot{\pi} + C(\tau)\,\pi + k^2\,D(\tau,k)\,\pi + H_0\,E(\tau,k) = 0,
 \label{eq:pi_eom}
\end{equation}
\end{widetext}
where the functions \(A\), \(B\), \(C\), \(D\), and \(E\) are defined in \cite{Hu:2014oga} and we report them in Appendix~\ref{app:pi_coeff}. 

We solve the system of equations comprising the Boltzmann equations Eqs.~\eqref{eq:photon1}–\eqref{eq:colddark2}, the modified Einstein equations Eqs.~\eqref{eq:hornd1}–\eqref{eq:hornd2}, and the $\pi$ field equation Eq.~\eqref{eq:pi_eom}, under the assumption of adiabatic perturbations discussed below, to determine the leading‐order ICs. Importantly, we included the full $\pi$ field equation of motions considering non-zero constant $\{\Omega, \gamma_{1,2,3}\}$, which is essential for obtaining the precise ICs. In contrast, the implementation in \texttt{EFTCAMB} assumes that the ICs are close to GR; it uses the standard GR ICs for metric and matter perturbations, and sets the $\pi$ field to the particular solution of its equation, effectively neglecting the contributions of the functions \(A\) and \(B\)  \cite{Hu:2014oga}.

\section{\label{sec:results}Results}

\subsection{\label{sec:IC_Adiabatic}Initial Conditions in the Adiabatic Modes}
We are interested in adiabatic ICs, for which every species $i$ satisfies,
\begin{equation}
  \frac{\delta P_i}{\delta\rho_i} \;=\;\frac{\dot{\bar{P}}_i}{\dot{\bar{\rho}}_i}\,.
  \label{eq:adiabatic_std}
\end{equation}
Under this condition, the density perturbations of different species are related via
\begin{equation}
  \frac{\delta\rho_i}{1+w_i} \;=\;\frac{\delta\rho_j}{1+w_j}\,.
  \label{eq:adiabatic_rel}
\end{equation}
Adiabatic conditions can be understood as generating from a homogeneous Universe to which a common, local shift in time, $\tau\rightarrow\tau+\delta\tau(\vec{x})$ has been applied to all background quantities. In \cite{Liu:2017oey}, the authors used this to set $\delta\tau = \pi$ as an adiabatic initial condition (IC) for the  St\"uckelberg field. However, the scalar field in EFT represents generically the DE or MG degree of freedom, and  is not necessarily in thermal equilibrium with the other species. In this work therefore, we do not set $\pi=\delta\tau$, but rather solve the equation for $\pi$ along with the other equations. This is one of the key differences in our implementation compared to \cite{Liu:2017oey}, which will allows us to find self-consistent solutions to the set of ICs equations.

Following the usual procedure, we set ICs on super horizon scales, i.e. for modes satisfying  $k\tau\ll1$, using $k\tau$ as a dimensionless expansion parameter. We expand all scalar perturbations as a Taylor series in the \(k\tau\) up to fourth order:

\begin{subequations}
\begin{align}
\delta_\alpha(k,\tau)
&= \sum_{n=0}^{4} \frac{X_{\alpha,n}}{n!}\,(k\tau)^n,\\
\theta_\alpha(k,\tau)
&= k \sum_{n=0}^{4} \frac{T_{\alpha,n}}{n!}\,(k\tau)^n,\\
\dot h(k,\tau)
&= k \sum_{n=0}^{4} \frac{H_{D,n}}{n!}\,(k\tau)^n,\\
\sigma_\nu(k,\tau)
&= \sum_{n=0}^{4} \frac{S_{n}}{n!}\,(k\tau)^n,\\
\eta(k,\tau)
&= \sum_{n=0}^{4} \frac{H_{E,n}}{n!}\,(k\tau)^n,\\
\pi(k,\tau)
&= \sum_{n=0}^{4} \frac{F_{P,n}}{n!}\,(k\tau)^n.
\end{align}\label{eq:IC_Adiabatic}
\end{subequations}
Here the species index \(\alpha\in\{b,c,\gamma,\nu\}\) labels baryons, cold dark matter, photons, and massless neutrinos, respectively. $X_{\alpha,n}$, $T_{\alpha,n}$, $H_{D,n}$, $S_n$, $H_{E,n}$, and $F_{P,n}$ are the coefficients of the expansions. Eventually we want to solve a system of algebraic equations at the lowest order in $k\tau$. In synchronous gauge, metric perturbation $h$ starts at order $\mathcal{O}(k^2\tau^2)$; thus, the Einstein and Boltzmann equations generate terms up to $\mathcal{O}(k^3\tau^3)$. Expanding to $\mathcal{O}(k^4\tau^4)$ order guarantees that all leading contributions to the ICs are accounted for.

Substituting these expansions into the set of Einstein–Boltzmann and $\pi$ field equations, and imposing the adiabatic condition for matter species, one obtains a set of algebraic relations at each order in \(k\tau\). Retaining only the first leading order in \(k\tau\) yields the leading‐order coefficients $X_{\alpha,n}$, $T_{\alpha,n}$, $H_{D,n}$, $S_n$, $H_{E,n}$, and $F_{P,n}$. The complete set of IC equations for the case of constant ${\Omega, \gamma_{1,2,3}}$ is provided in Appendix~\ref{app:full_ICs}. For the specific case of $\gamma_{1,2,3}=0$ with \(\Omega=\mathrm{const}\), the leading‐order ICs are the following:

\begin{align}
h &= \frac{5 k^2 \tau^2 (1 + \Omega)}{10 + 6 \Omega}, \\
\delta_b &=\delta_c= -\frac{5 k^2 \tau^2 (1 + \Omega)}{4 (5 + 3 \Omega)}, \\
\delta_\gamma &=\delta_\nu= -\frac{5 k^2 \tau^2 (1 + \Omega)}{15 + 9 \Omega}, \\
\theta_c &= 0, \theta_\gamma =\theta_b = -\frac{5 k^4 \tau^3 (1 + \Omega)}{36 (5 + 3 \Omega)},\\
\theta_\nu &= -\frac{k^4 \tau^3 (1 + \Omega) (115 + 99 \Omega + 20 R_\nu)}{36 (5 + 3 \Omega) (15 (1 + \Omega) + 4 R_\nu)}, \\
\sigma_\nu &= \frac{2 k^2 \tau^2 (1 + \Omega)}{45 (1 + \Omega) + 12 R_\nu}, \\
\eta &= 1 - \frac{5 k^2 \tau^2 (1 + \Omega) (5 + 9 \Omega + 4 R_\nu)}{12 (5 + 3 \Omega) (15 (1 + \Omega) + 4 R_\nu)}\,,
\end{align}
where $R_\nu\equiv\bar{\rho}_\nu/(\bar{\rho}_\nu+\bar{\rho}_\gamma)$.
In addition, the solution for the \(\pi\) field is given by:
\begin{equation}
\pi(k, \tau) = -\frac{k^2 \tau^3 H_0 (1 + \Omega_0)}{4 (5 + 3 \Omega_0)}\,.
\end{equation}
We emphasize that these solutions differ from those reported in \cite{Liu:2017oey}, because of the key difference in our treatment of $\pi$, which is not set to correspond to $\delta \tau$, as discussed before.

For comparison, \texttt{hi\_class} reported ICs for a subcase of kinetic gravity braiding (KGB) models, where the $\mathrm{\alpha}$ parameters are constant:
\begin{equation}
\mathrm{\Omega_{DE}},\;\; \mathrm{\alpha_K},\;\; \mathrm{\alpha_B} = \mathrm{const}, \qquad \mathrm{\alpha_M} = \mathrm{\alpha_T} = 0, \qquad \mathrm{M^2_*} = 1\,,
\end{equation}
as shown in \cite{Bellini:2019syt}. See the definitions of these parameters ($\mathrm{\Omega_{DE}}$, $\mathrm{\alpha_K}$, $\mathrm{\alpha_B}$, $\mathrm{\alpha_M}$, $\mathrm{\alpha_T}$, and $\mathrm{M^2_*}$) in Section~2 of \cite{Bellini:2019syt}. After translating alphas and $\Omega_{\mathrm{DE}}$ into the EFT parameter $\Omega$, we have explicitly verified that our ICs with $\Omega=\mathrm{const}$ and $\gamma_{1,2,3}=0$ reproduce the KGB sub-case solutions derived in \cite{Bellini:2019syt} (see their Eqs.~(4.12)–(4.13)) in the case of a perfect-fluid dark energy with constant parameter $w$ and sound speed $c_s$.

In the following section, we demonstrate that our ICs produce the correct behaviors of both $\delta_\gamma$ and the Weyl potential $(\Psi+\Phi)$ in the MG models, as opposed to the approximated ICs derived from GR.

\begin{figure}[h]
\includegraphics[width=1\linewidth]{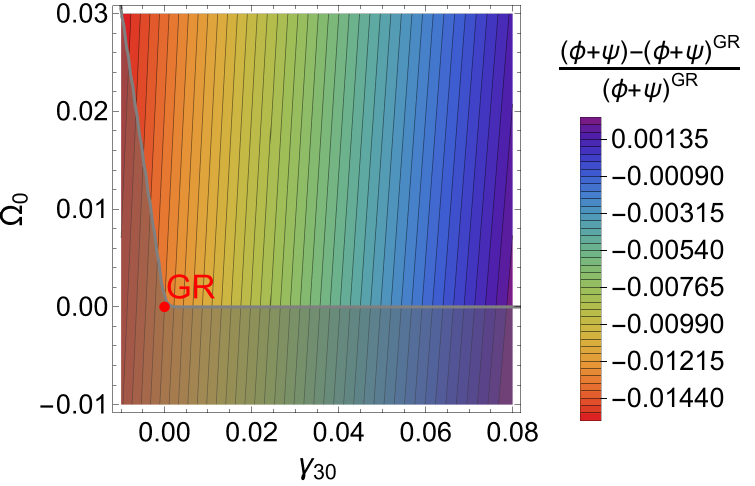}
\caption{Fractional difference in the MG Weyl potential relative to its GR initial condition, i.e., $((\Psi+\Phi)-(\Psi+\Phi)^{\mathrm{GR}})/(\Psi+\Phi)^{\mathrm{GR}}$. The figure is color-coded according to this ratio for different values of $\Omega_0$ and $\gamma_{30}$, with $\gamma_{20}$ fixed at zero. Adopting consistent MG ICs can produce $\sim1.4\%$ differences in the Weyl potential during the radiation-dominated epoch. The red dot indicates the GR value. The grey region indicates parameter space excluded by stability conditions: ghost instability (negative kinetic term) or gradient instability (sound speed $c_s^2 < 0$).
}
\label{fig:curvature_dff}
\end{figure}

Before proceeding with a more detailed analysis, we first examine the influence of MG ICs on the evolution of perturbations, focusing specifically on the Weyl potential. The metric perturbations in the conformal Newtonian gauge is given by
\begin{equation}
ds^2 = a^2(\tau)\left[-(1+2\Psi)\,d\tau^2 + (1-2\Phi)\,dx^i dx_i\right]
\label{eq:metric_newtonian}
\end{equation}
with $\Psi$ denoting the Newtonian potential and $\Phi$ denoting the intrinsic spatial curvature
potential, as defined in \cite{Hu:1994uz,Hu:1997mn}. In the radiation-dominated era, and under the assumption $\gamma_{20}=0$ (which is justified since  $\gamma_{20}=0$ contributes only at the next-to-leading order in the MG ICs), the analytical expression for the Weyl potential $(\Psi+\Phi)$ is given by
\begin{equation}
\Psi+\Phi = \frac{
-4R_{\nu} - 5(4+3\gamma_{30}+4\Omega_0)
}{
4R_{\nu} + 15\left(1 + \gamma_{30} + \Omega_0\right)
}.
\end{equation} 
The Weyl potential influences gravitational lensing and the integrated Sachs–Wolfe effect, and its drop after horizon crossing sources the photon acoustic oscillations \cite{Hu:1994uz,Lin:2018nxe}
\begin{eqnarray}
    &&[\Theta_0-\Phi](\tau) = [\Theta_0-\Phi](0)\cos(kc_s\tau)\nonumber\\
        &&\qquad -\frac{k}{\sqrt{3}}\int_0^\tau d\tau'[\Psi+\Phi](\tau')\sin[kc_s(\tau-\tau')],
\end{eqnarray}
where $\Theta_0=\frac{1}{4}\delta_\gamma$ is the photon monopole perturbation, and $c_s=1/\sqrt{3}$ is the sound speed of photon.
All of these are important for the modified CMB angular power spectrum, which we will discuss in \cref{sec:Cosmo_IC}.
\cref{fig:curvature_dff} shows the fractional difference between the Weyl potential calculated with consistent MG ICs and that evolved with GR ICs, evaluated in the radiation era. As demonstrated, variations in \(\gamma_{30}\) have a more pronounced impact on the evolution of Weyl potential than $\Omega_0$.

We also observe that the IC for the Weyl potential can diverge when its denominator vanishes. This divergence occurs in the parameter space which is unstable according to ghost and gradient stability conditions (indicated by the gray-shaded areas in \cref{fig:curvature_dff}). Therefore, the divergence in Weyl potential does not affect cosmological parameter inference, since such pathological behavior happens in the theoretically excluded parameter space. Because we have only considered constant EFT functions at early times, the results here only apply for EFT functions behaving as constant functions. Whether other choices of EFT functions that satisfy the stability conditions may still encounter regions with divergent ICs needs to be examined separately.

The results indicate that the effects of using consistent MG ICs become prominent in certain regions of the EFT parameter space. In \cref{sec:Cosmo_IC}, we demonstrate that such differences translate into non-negligible modifications in the CMB angular spectra exploring several representative  choices of \(\Omega_0\) and \(\gamma_{30}\).

\begin{figure*}[]
\includegraphics[width=1\linewidth]{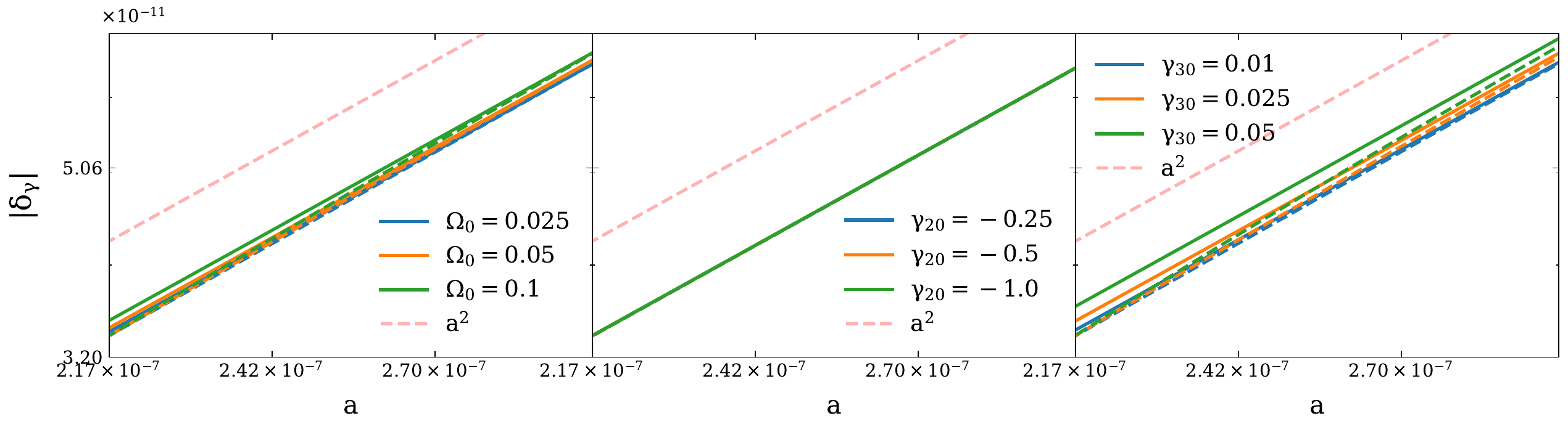}
\caption{Evolution of the absolute photon density perturbation, \( |\delta_\gamma(a)| \), for MG models. From left to right, the panels vary the EFT functions \(\Omega_0\), \(\gamma_{20}\), and \(\gamma_{30}\), respectively, while the remaining functions are fixed at their baseline values (\(\Omega_0 = 0.001\) and \(\gamma_{10} = \gamma_{20} = \gamma_{30} = 0\)). All cases use identical MG equations of motion and differ only in their ICs. Solid lines correspond to the evolution using the correct MG ICs, whereas the dashed lines represent the evolution with GR ICs. An \(a^2\) scaling function is plotted to represent the expected \(\delta_\gamma\) evolution, which agrees with the  \( |\delta_\gamma(a)| \) evolution using MG ICs.}
\label{fig:delta_photons}
\end{figure*}

\subsection{Validity of Modified Gravity Initial Conditions}

Figure~\ref{fig:delta_photons} presents the evolution of the photon density perturbation, \( |\delta_\gamma(a)| \),  with $k=1\times10^{-4}$ Mpc$^{-1}$ in MG models. We vary \(\Omega_0\), \(\gamma_{20}\), and \(\gamma_{30}\) one at a time, while keeping the other EFT functions fixed at their baseline values:  \(\Omega_0 = 0.001\) and \(\gamma_{10} = \gamma_{20} = \gamma_{30} = 0\). By choosing a small value for $\Omega_0$, we ensure that the stability conditions are satisfied, and the effects of each EFT function on the evolution of the perturbation can be isolated.

Note that the MG equations of motion are the same for all cases , and only the ICs differ. It can be seen that the evolution of $\delta_\gamma$ with GR ICs fail to reproduce the expected $\delta_\gamma$ power-law evolution at the beginning, and it converges to the correct evolution with MG ICs with the proper \(a^2\) scaling later. This underscores the need for consistent MG ICs in early MG models. Note that the evolution of $\delta_\gamma$  does not change with \(\gamma_{20}\) as expected as \(\gamma_{20}\) only affects the next-to-leading order term of $\delta_\gamma$.

\begin{figure*}[]
\includegraphics[width=\linewidth]{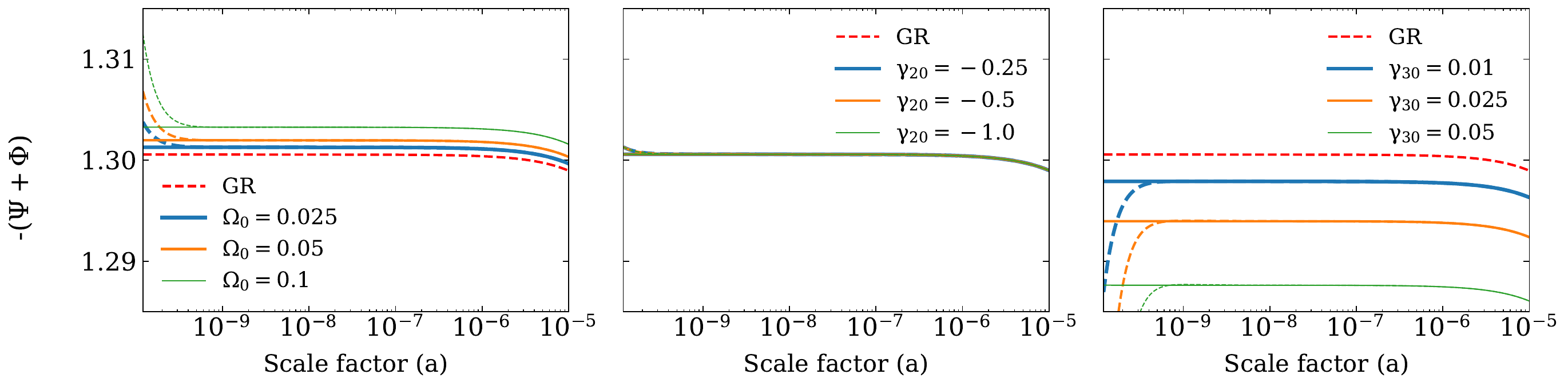}
\caption{Evolution of the Weyl potential, $-(\Psi + \Phi)$, with varying \(\Omega_0\), \(\gamma_{20}\), and \(\gamma_{30}\), respectively. The remaining EFT functions remain fixed at their baseline values (\(\Omega_0 = 0.001\) and \(\gamma_{10} = \gamma_{20} = \gamma_{30} = 0\)). Solid lines (using MG ICs) demonstrate that $-(\Psi + \Phi)$ is conserved at early times, whereas dashed lines (GR ICs) show significant deviations from it. }
\label{fig:curvature}
\end{figure*}

\Cref{fig:curvature} illustrates the evolution of the Weyl potential $(\Psi + \Phi)$. For clarity of visualization, we plot $-(\Psi + \Phi)$. Each subplot independently varies one of the EFT functions $\{\Omega_0$, $\gamma_{20}$, or $\gamma_{30}\}$, while keeping the remaining parameters fixed at the baseline values mentioned above. Notably, when using the consistent MG ICs (solid lines), Weyl potential is conserved at early times. In contrast, the GR ICs (dashed lines) results in significant deviations from the expected Weyl potential values, demonstrating the non-conservation of Weyl potential under inconsistent GR ICs. Again $\gamma_{20}$ has little effect on the Weyl potential. The observed difference in Weyl potential between MG and GR ICs when varying $\gamma_{20}$ arises primarily due to the choice of a fixed $\Omega_0 = 0.01$. This difference diminishes as the value of $\Omega_0$ decreases. These results further validates our ICs solutions, ensuring conservation of Weyl potential in early-time MG models at super-horizon scale.

\subsection{\label{sec:Cosmo_IC}Impact on Cosmological Observables}

\begin{figure*}[]
\centering
\includegraphics[width=\linewidth]{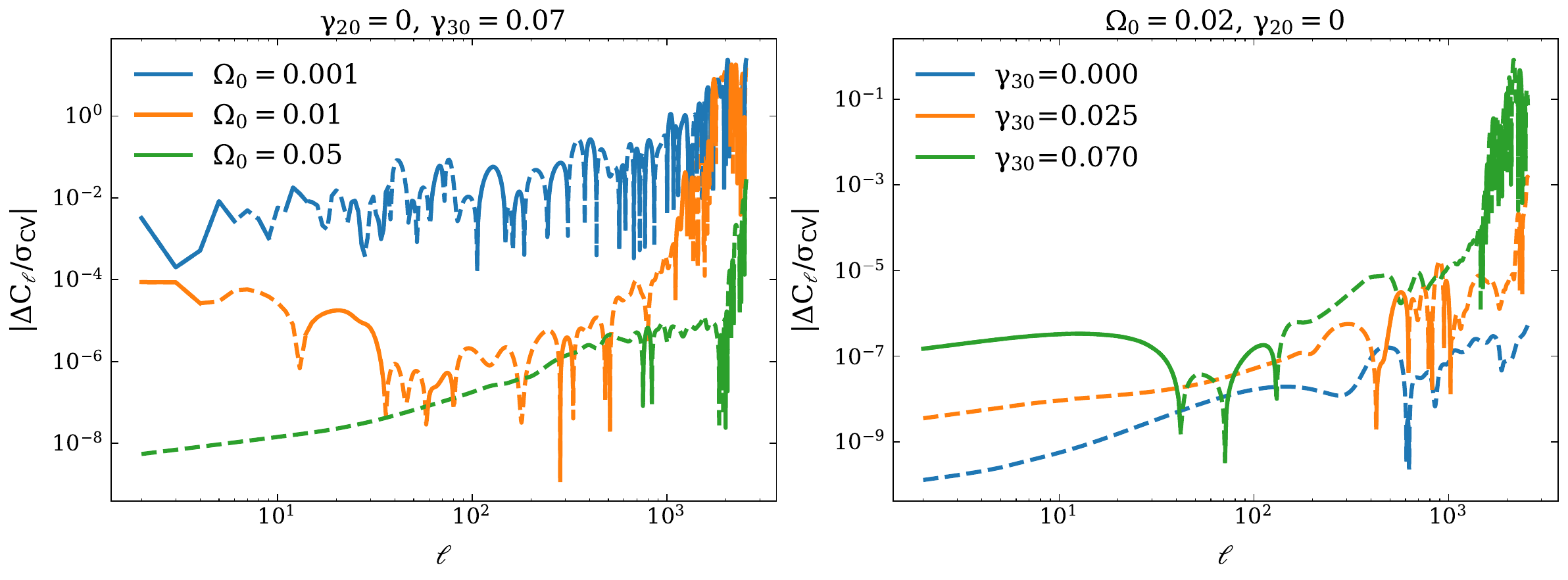}
\caption{Absolute value of the relative difference in the CMB TT angular power spectrum, \(|\Delta C_\ell/\sigma_{\mathrm{CV}}| = |(C_{\ell,\text{MG ICs}} - C_{\ell,\text{GR ICs}})/\sigma_{\mathrm{CV}}|\), for various MG models specified by \(\Omega_0\), \(\gamma_{20}\), and \(\gamma_{30}\). Solid lines indicate positive relative differences, and dashed lines are negative differences. In the left panel, $\gamma_{20}=0$ and $\gamma_{30}=0.07$ are held fixed, while $\Omega_0$ is varied; these choices lie close to the instability boundary, and the difference between the GR and MG ICs is small. In the right panel, $\Omega_0=0.02$ and $\gamma_{20}=0$ are fixed, while $\gamma_{30}$ is varied; these choices stay away from the instability boundary, and have a large discrepancy between the GR and MG ICs. $C_{\ell,\text{MG ICs}}$ and $C_{\ell,\text{GR ICs}}$ only differ in using the consistent MG ICs and the inconsistent GR ICs. The relative difference is scaled with cosmic variance per multiple of the $C_{\ell,\text{MG ICs}}$.}
\label{fig:EFTCL}
\end{figure*}

Variations in the EFT functions can alter either the expansion history or the evolution of perturbations, thereby affect the CMB angular power spectrum. Specifically, $\Omega$ modifies both the background expansion and perturbation evolution, while $\gamma_{1}$, $\gamma_{2}$, and $\gamma_{3}$ affect the evolution of perturbations. To illustrate the importance of correctly incorporating MG ICs in MG models, we compare the CMB temperature angular power spectra \(C_{\ell}\) in MG models obtained using two different ICs. \cref{fig:EFTCL} shows the relative difference,
$|\Delta C_\ell| \equiv |C_{\ell,\text{MG ICs}}-C_{\ell,\text{GR ICs}}|$, where \(C_{\ell,\text{MG ICs}}\) uses the correct MG ICs, while the \(C_{\ell,\text{GR ICs}}\) is computed using the GR ICs. We scaled $|\Delta C^{TT}_\ell|$ with the cosmic variance per multipole defined as the following.
\begin{equation}
\sigma_{\mathrm{CV}} = \sqrt{\frac{2}{2\ell+1}}\; C_{\ell,\mathrm{MG\;ICs}}.
\end{equation}

In the left panel of \cref{fig:EFTCL}, we fix $\gamma_{20}=0$ and $\gamma_{30}=0.07$, and vary $\Omega_0$ over the values 0.001, 0.01, and 0.05. Here solid lines correspond to positive relative differences $\Delta C_\ell/\sigma_{\mathrm{CV}}$, and dashed lines are the negative relative differences plotted as $-\Delta C_\ell/\sigma_{\mathrm{CV}}$. The relative difference $|\Delta C_\ell/\sigma_{\mathrm{CV}}|$ can exceed 0.1 and reach values as large as $\sim10$ for multipoles $\ell > 1000$, with smaller $\Omega_0$ leading to larger deviations. This effect is becoming larger as $\Omega_0$ decreases and the model approaches the instability boundary in \cref{fig:curvature_dff}, defined by the ghost and gradient stability conditions. In this process, although the difference between the GR and MG ICs stays similar as shown in \cref{fig:delta_photons}, the proximity to the instability boundary dynamically amplifies its effect during the subsequent evolution, resulting in significant deviations in $C_\ell$. For example, with $\Omega_0=0.001$, $|\Delta C_\ell|$ can exceed $\mathcal{O}(10\%)$ over the range $10\lesssim \ell \lesssim 2000$. This underscores the importance of using consistent MG ICs, especially when the EFT parameters place the model near a stability boundary.

In the right panel of \cref{fig:EFTCL}, we fix $\Omega_{0}=0.02$ and $\gamma_{20}=0$, and vary $\gamma_{30}$ over the values 0.0, 0.025, and 0.07. In this case, the EFT functions are far away from the unstable region, yet there is a large increase of the Weyl potential calculated with the MG ICs (see \cref{fig:curvature_dff}). The relative difference
$|\Delta C_\ell/\sigma_{\mathrm{CV}}|$ still becomes non-negligible as $\gamma_{30}$ becomes 0.07 for multipoles $\ell > 1000$, which is mainly driven by the modifications to the ICs.

The deviations shown in \cref{fig:EFTCL} can also be interpreted relative to the cosmic-variance limits of real experiments. Planck is effectively cosmic-variance limited in temperature up to multipoles $\ell \sim 1500$, with fractional uncertainties of order a few percent at $\ell \sim 1000$ \citep{Planck:2019nip}. The Simons Observatory and proposed CMB-S4 surveys, with $f_{\rm sky}\sim0.4$ and noise levels of a few $\mu{\rm K}$-arcmin, are forecast to remain close to the cosmic-variance limit over a similar range of scales \citep{SimonsObservatory:2018koc,CMB-S4:2016ple}. Comparing with our results, we find that $|\Delta C_\ell|/\sigma_{\mathrm{CV}}$ can exceed unity for $\ell \gtrsim 1000$ in several MG parameter choices, indicating that these effects are, in principle, observable with Planck-quality data and even more readily detectable by SO- or CMB-S4-like surveys.

To further assess the impact on cosmological inference, we propagate the \(C_\ell\) shifts through the Fisher bias formalism (e.g. \citep{Knox:1998fp,Huterer:2000mj}). For an observed data vector \(D_\alpha\) (here, the \(C_\ell\)’s) with covariance \(C\), the first-order bias in parameter \(p_i\) due to a shift \(\Delta D_\beta\) is
\begin{equation}
\Delta p_i \;=\; \sum_j (F^{-1})_{ij}\;\sum_{\alpha\beta}\;
\frac{\partial \bar D_\alpha}{\partial p_j}\,
(C^{-1})_{\alpha\beta}\,\Delta D_\beta ,
\label{eq:bias}
\end{equation}
where \(F\) is the Fisher matrix. We assume a CMB survey with sky coverage \(f_{\rm sky}=0.48\) (SO-like)
temperature noise of a few \(\mu\mathrm{K}\)-arcmin, and a Gaussian beam with arcminute-scale resolution, so that the covariance is Gaussian with 
\(\mathrm{Var}(C_\ell)=\tfrac{2}{(2\ell+1)f_{\rm sky}}(C_\ell+N_\ell)^2\). 
For the left panel case in \cref{fig:EFTCL} with \(\Omega_0=0.001, \gamma_{20}=0, \gamma_{30}=0.07\), the resulting biases in units of the \(1\sigma\) Fisher uncertainties are 
\(\Delta H_0/\sigma\simeq -0.52,\; 
\Delta \sigma_8/\sigma\simeq 0.038,\; 
\Delta \omega_b/\sigma\simeq -0.62,\; 
\Delta \omega_c/\sigma\simeq 0.24,\; 
\Delta n_s/\sigma\simeq 0.10\), 
indicating that ignoring consistent MG initial conditions can induce \(\mathcal{O}(0.2\text{--}0.5)\sigma\) shifts in several cosmological parameters.
 
The oscillatory features (the "wiggles") in the curves of \cref{fig:EFTCL} stem from the interpolation used in the \texttt{EFTCAMB}, and the overall trend reflects the impact of incorporating the correct MG ICs. Therefore, incorporating correct MG ICs can produce measurable effects in the CMB angular power spectrum with early MG models, where both stability conditions and corrections in ICs play important roles. This emphasizes the need for accurate ICs in early MG models.

\section{Conclusion}

In this work, we derived the consistent ICs of MG models using EFT approach, specifically covering the Horndeski theory parameter space. This is a key progress towards the consistent modeling of DE/MG models starting at the early universe in the Einstein-Boltzmann solver. We found that incorporating the correct MG ICs leads to non-negligible corrections in the early-universe cosmology in MG models compared to the standard GR ICs.

 Using non-zero constant EFT functions \(\Omega_0\) and \(\gamma_{1,2,3}\), we demonstrate our derived ICs correctly reproduce the evolution of both photon density perturbations and the Weyl potential conservation. In contrast, evolution using the standard GR ICs can lead to significant deviations of the correct $\delta_\gamma$ and Weyl potential in MG models at the early time.

 The discrepancy in MG evolutions from using incorrect ICs can yield measurable changes in the theoretical modeling of cosmological observables. We found that the CMB angular power spectrum—an important probe of MG effects e.g., \cite{Planck:2015bue,ACT:2025tim}—can exhibit differences of up to \(\sim10\,\sigma_{\mathrm{CV}}\) when incorrect ICs are used. Moreover, Fisher forecasts indicate that such differences can bias cosmological parameters at the level of \(\mathcal{O}(0.2\text{--}0.5)\sigma\), further emphasizing the importance of consistent MG ICs.

Our findings underscore the need to adopt consistent MG ICs to place reliable constraints on MG models. 
Our IC solutions apply to the case where the EFT functions are approximately constant in the early Universe when ICs are set up in the code. If the EFT functions have strong time-dependent evolution when ICs are set up, our solutions may break down, and we leave the generalization for future work. Note that commonly used EFT functions e.g.,\ $\Omega(a)\propto a^n$ or $\Omega(a)\propto \Omega_{\rm DE}$, would be very small in the deep radiation era, so their ICs would cause very small effects on the cosmological observables. Also, it is possible to extend the MG ICs to the isocurvature modes. 

\begin{acknowledgments}
JP acknowledges support from the Leinweber Center for Theoretical Physics and DOE under contract DE-SC009193.
M-X.L. is supported by funds provided by the Center for Particle Cosmology.
AS acknowledges support from the European Research Council under the H2020 ERC Consolidator Grant “Gravitational Physics from the Universe Large scales Evolution” (Grant No. 101126217 — GraviPULSE) and from the NWO and the Dutch Ministry of Education, Culture and Science (OCW) (through NWO VIDI Grant No. 2019/ENW/00678104).
\end{acknowledgments}

\appendix
\section{Full Equations for Initial Conditions \label{app:full_ICs}}
Here we present the full ICs for non-zero constant $\{\Omega, \gamma_{1,2,3}\}$ of Horndeski models in the EFT approach:
\begin{subequations}\label{eq:full_ics}
\begin{widetext}
\begin{align}
\dot{h} &= \frac{k^2 \tau}{40(1+\gamma_{30}+\Omega_0)(3\gamma_{30}(1+\Omega_0)+\Omega_0(5+3\Omega_0))
\Bigl(2\Omega_0(4+3\Omega_0)+\gamma_{30}(3+6\Omega_0)\Bigr)} \label{eq:hdot}
\\ &\quad \times \Bigl[-27\gamma_{30}^4\Bigl(-20+9\tau H_0\gamma_{20}-40\Omega_0\Bigr) \nonumber
\\ &\quad \quad +\, 4\Omega_0^2(1+\Omega_0)\Bigl(400+(700+27\tau H_0\gamma_{20})\Omega_0+300\Omega_0^2\Bigr) \nonumber
\\ &\quad \quad +\, 15\gamma_{30}^2\Bigl(24+(332-18\tau H_0\gamma_{20})\Omega_0+(724-9\tau H_0\gamma_{20})\Omega_0^2+392\Omega_0^3\Bigr) \nonumber
\\ &\quad \quad +\, 2\gamma_{30}\Omega_0\Bigl(780+(4040+81\tau H_0\gamma_{20})\Omega_0+4(1340+27\tau H_0\gamma_{20})\Omega_0^2+2100\Omega_0^3\Bigr) \nonumber
\\ &\quad \quad -\, 18\gamma_{30}^3\Bigl(9\tau H_0\gamma_{20}(2+3\Omega_0)-10(5+26\Omega_0+22\Omega_0^2)\Bigr)\Bigr], \nonumber \\[2ex]
\delta_b &= \delta_c = -\frac{k^2 \tau^2 \Bigl(6\gamma_{30} + 9\gamma_{30}^2 + 10\Omega_0 + 15\gamma_{30}\Omega_0 + 10\Omega_0^2\Bigr)}
{8 \Bigl(3\gamma_{30} + 5\Omega_0 + 3\gamma_{30}\Omega_0 + 3\Omega_0^2\Bigr)} \label{eq:deltab}
\\[1ex] &\quad + \frac{9k^2 \tau^3 H_0 (3\gamma_{30} - 2\Omega_0)(\gamma_{20}\gamma_{30} + \gamma_{20}\Omega_0)(4\gamma_{30} + 3\gamma_{30}^2 + 2\Omega_0 + 5\gamma_{30}\Omega_0 + 2\Omega_0^2)}
{80 (1+\gamma_{30}+\Omega_0)(3\gamma_{30}+5\Omega_0+3\gamma_{30}\Omega_0+3\Omega_0^2)(3\gamma_{30}+8\Omega_0+6\gamma_{30}\Omega_0+6\Omega_0^2)}, \nonumber \\[2ex]
\delta_\gamma &= \delta_\nu = -\frac{k^2 \tau^2 \Bigl(6\gamma_{30} + 9\gamma_{30}^2 + 10\Omega_0 + 15\gamma_{30}\Omega_0 + 10\Omega_0^2\Bigr)}
{6 \Bigl(3\gamma_{30}+5\Omega_0+3\gamma_{30}\Omega_0+3\Omega_0^2\Bigr)} \label{eq:deltagamma}
\\[1ex] &\quad + \frac{3k^2 \tau^3 H_0 (3\gamma_{30}-2\Omega_0)(\gamma_{20}\gamma_{30}+\gamma_{20}\Omega_0)(4\gamma_{30}+3\gamma_{30}^2+2\Omega_0+5\gamma_{30}\Omega_0+2\Omega_0^2)}
{20 (1+\gamma_{30}+\Omega_0)(3\gamma_{30}+5\Omega_0+3\gamma_{30}\Omega_0+3\Omega_0^2)(3\gamma_{30}+8\Omega_0+6\gamma_{30}\Omega_0+6\Omega_0^2)}, \nonumber \\[2ex]
\theta_c &= 0, \theta_\gamma =\theta_b = -\frac{k^4 \tau^3 \Bigl(6\gamma_{30}+9\gamma_{30}^2+10\Omega_0+15\gamma_{30}\Omega_0+10\Omega_0^2\Bigr)}
{72 \Bigl(3\gamma_{30}+5\Omega_0+3\gamma_{30}\Omega_0+3\Omega_0^2\Bigr)}, \label{eq:thetab} \\[2ex]
\theta_\nu &=  
-\frac{k^4 \tau^3 \Bigl(135\gamma_{30}^3 + 9\gamma_{30}^2(25 + 4R_\nu + 40\Omega_0) + 2\Omega_0(1+\Omega_0)(115 + 20R_\nu + 99\Omega_0) + 3\gamma_{30}(46 + 187\Omega_0 + 141\Omega_0^2 + 4R_\nu(2+5\Omega_0))\Bigr)}
{72 \Bigl(4R_\nu + 15(1+\gamma_{30}+\Omega_0)\Bigr)\Bigl(3\gamma_{30}(1+\Omega_0)+\Omega_0(5+3\Omega_0)\Bigr)} \label{eq:thetanu}
\\[2ex] &\quad + \frac{3k^4 \tau^4 H_0\gamma_{20}(\gamma_{30}+\Omega_0)
\Bigl(3\gamma_{30}^2+2\Omega_0(1+\Omega_0)+\gamma_{30}(4+5\Omega_0)\Bigr)
\Bigl(45\gamma_{30}^2+3\gamma_{30}(5+2R_\nu-15\Omega_0)-2\Omega_0(55+2R_\nu+45\Omega_0)\Bigr)}
{320 \Bigl(1+\gamma_{30}+\Omega_0\Bigr)\Bigl(2R_\nu+15(1+\gamma_{30}+\Omega_0)\Bigr)\Bigl(3\gamma_{30}(1+\Omega_0)+\Omega_0(5+3\Omega_0)\Bigr)\Bigl(2\Omega_0(4+3\Omega_0)+\gamma_{30}(3+6\Omega_0)\Bigr)}, \nonumber \\[2ex]
\eta &= 1 - \frac{k^2 \tau^2 \Bigl(135\gamma_{30}^3+10\Omega_0(1+\Omega_0)(5+4R_\nu+9\Omega_0)+9\gamma_{30}^2(25+4R_\nu+40\Omega_0)
+3\gamma_{30}\Bigl(4R_\nu(2+5\Omega_0)+5(2+23\Omega_0+21\Omega_0^2)\Bigr)\Bigr)}
{24\Bigl(4R_\nu+15(1+\gamma_{30}+\Omega_0)\Bigr)\Bigl(3\gamma_{30}(1+\Omega_0)+\Omega_0(5+3\Omega_0)\Bigr)} \label{eq:eta}
\\ &\quad + \frac{3k^2 \tau^3 H_0\gamma_{20}(\gamma_{30}+\Omega_0)
\Bigl(3\gamma_{30}^2+2\Omega_0(1+\Omega_0)+\gamma_{30}(4+5\Omega_0)\Bigr)
\Bigl(90\gamma_{30}^2+2\Omega_0(70-4R_\nu+45\Omega_0)+3\gamma_{30}(55+4R_\nu+60\Omega_0)\Bigr)}
{160\Bigl(1+\gamma_{30}+\Omega_0\Bigr)\Bigl(2R_\nu+15(1+\gamma_{30}+\Omega_0)\Bigr)\Bigl(3\gamma_{30}(1+\Omega_0)+\Omega_0(5+3\Omega_0)\Bigr)\Bigl(2\Omega_0(4+3\Omega_0)+\gamma_{30}(3+6\Omega_0)\Bigr)}, \nonumber \\[2ex]
\sigma_\nu &= \frac{2k^2 \tau^2 \Bigl(3\gamma_{30}+5\Omega_0+6\gamma_{30}\Omega_0+8\Omega_0^2+3\gamma_{30}\Omega_0^2+3\Omega_0^3\Bigr)}
{3\Bigl(15+4R_\nu+15\gamma_{30}+15\Omega_0\Bigr)\Bigl(3\gamma_{30}+5\Omega_0+3\gamma_{30}\Omega_0+3\Omega_0^2\Bigr)} \label{eq:sigmanu}
\\[1ex] &\quad + \frac{3k^2 \tau^3 (H_0\gamma_{20}\gamma_{30}+H_0\gamma_{20}\Omega_0)
(4\gamma_{30}+3\gamma_{30}^2+2\Omega_0+5\gamma_{30}\Omega_0+2\Omega_0^2)}
{8\Bigl(1+\gamma_{30}+\Omega_0\Bigr)\Bigl(15+2R_\nu+15\gamma_{30}+15\Omega_0\Bigr)\Bigl(3\gamma_{30}+5\Omega_0+3\gamma_{30}\Omega_0+3\Omega_0^2\Bigr)}. \nonumber
\end{align}
\end{widetext}
\end{subequations}
The equation for the $\pi$ field is the following.
\begin{widetext}
\begin{equation}
\begin{split}
\pi =\;& -\frac{k^2 \tau^3 H_0 \Bigl(3\gamma_{30}^2+2\Omega_0(1+\Omega_0)+\gamma_{30}(4+5\Omega_0)\Bigr)}
{8\Bigl(3\gamma_{30}(1+\Omega_0)+\Omega_0(5+3\Omega_0)\Bigr)}\\[1ex]
&+\frac{9\,k^2 \tau^4 H_0^2\Bigl(5+3\gamma_{30}+3\Omega_0\Bigr)
\gamma_{20}(\gamma_{30}+\Omega_0)
\Bigl(4\gamma_{30}+3\gamma_{30}^2+2\Omega_0+5\gamma_{30}\Omega_0+2\Omega_0^2\Bigr)}
{160(1+\gamma_{30}+\Omega_0)
\Bigl(3\gamma_{30}+5\Omega_0+3\gamma_{30}\Omega_0+3\Omega_0^2\Bigr)
\Bigl(3\gamma_{30}+8\Omega_0+6\gamma_{30}\Omega_0+6\Omega_0^2\Bigr)}.
\label{eq:pi}
\end{split}
\end{equation}
\end{widetext}

\section{Modified Einstein Equations Coefficients \label{app:constEFT}}
Assuming EFT functions $\{\Omega, \gamma_{1,2,3}\}$ are constant, the six background operators for the Einstein equations in Horndeski theory, which indicates $2\gamma_5 = \gamma_3 = -\gamma_4$ and $\gamma_6 = 0$, are the following:
\vspace{-2ex}
\begin{widetext}
\begin{subequations}
\begin{align}
X &= 1 - \frac{\gamma_{40}}{1+\Omega_0}\,,\\[6pt]
F &= \frac{3a^2}{2k(1+\Omega_0)m_0^2}(\rho_Q+P_Q)\,\pi
       + \frac{3aH_0\gamma_{20}}{2k(1+\Omega_0)}(\dot\pi+\mathcal{H}\pi)\notag\\
     &\quad + \frac{3\gamma_{30}}{2(1+\Omega_0)}\Bigl[k - \frac{\dot{\mathcal{H}}-\mathcal{H}^2}{k}\Bigr]\pi
       + \frac{3\gamma_{40}}{2(1+\Omega_0)}\Bigl(k - \frac{\dot{\mathcal{H}}-\mathcal{H}^2}{k}\Bigr)\pi\,,\\[6pt]
U &= 1 + \frac{3\gamma_{30}}{2(1+\Omega_0)} + \frac{\gamma_{40}}{2(1+\Omega_0)}\,,\\[6pt]
G &= 1 + \frac{aH_0\gamma_{20}}{2\mathcal{H}(1+\Omega_0)}
         + \frac{3\gamma_{30}}{2(1+\Omega_0)} + \frac{\gamma_{40}}{2(1+\Omega_0)}\,,\\[6pt]
\mathcal{Q} &= 1 + \frac{2\gamma_{50}}{1+\Omega_0}\,,\\[6pt]
L &= \frac{a^2\dot\rho_Q}{2\mathcal{H}(1+\Omega_0)m_0^2}\,\pi
       + \frac{a^2c}{\mathcal{H}(1+\Omega_0)m_0^2}(\dot\pi+\mathcal{H}\pi)
       + \frac{2a^2H_0^2\gamma_{10}}{\mathcal{H}(1+\Omega_0)}(\dot\pi+\mathcal{H}\pi)\notag\\
     &\quad + \frac{3aH_0\gamma_{20}}{2(1+\Omega_0)}\Bigl(\frac{\dot{\mathcal{H}}}{\mathcal{H}}-2\mathcal{H}-\frac{k^2}{3\mathcal{H}}\Bigr)\pi
       - \frac{3aH_0\gamma_{20}}{2(1+\Omega_0)}\dot\pi\notag\\
     &\quad - \frac{3\gamma_{30}}{2(1+\Omega_0)}\bigl[k^2 - 3(\dot{\mathcal{H}}-\mathcal{H}^2)\bigr]\pi
       + \frac{3\gamma_{40}}{2(1+\Omega_0)}\Bigl[\dot{\mathcal{H}}-\mathcal{H}^2-\tfrac{k^2}{3}\Bigr]\pi\notag\\
     &\quad + \frac{2\gamma_{50}}{1+\Omega_0}\,k^2\pi\,.
\end{align}
\end{subequations}
\end{widetext}
\vspace{-2ex}

Here the quantities (\(\rho_Q,P_Q\)) are energy density and pressure of the effective EFT dark fluid \cite{Hu2014}. 
\begin{align}
\rho_Q&= 2c-\Lambda\,,\\
P_Q&= \Lambda\,.
\end{align}
The functions $c$ and $\Lambda$ are determined once the background expansion history is specified by the dark energy equation of state $w_{\rm DE}$. They are defined as,
\begin{align}
\Lambda &= -\frac{\Omega_0}{a^2}\bigl(2\dot{\mathcal{H}}+\mathcal{H}^2\bigr)
            + w_{\rm DE}\,\rho_{\rm DE}\,,\\
c       &= \frac{\Omega_0\bigl(\mathcal{H}^2 - \dot{\mathcal{H}}\bigr)}{a^2}
            + \tfrac{1}{2}\,\rho_{\rm DE}\,\bigl(1+w_{\rm DE}\bigr)\,.
\end{align}

\begin{widetext}
\section{\(\pi\) field Equation of Motion Coefficients \label{app:pi_coeff}}

Here we present coefficients \(A\), \(B\), \(C\), \(D\), and \(E\) for \(\pi\) field equation of motion assuming constant $\{\Omega, \gamma_{1,2,3}\}$ in the Horndeski models:
\begin{subequations}
\begin{align}
A\;=\;&
\frac{c\,a^2}{m_0^2}
\;+\;2\,a^2\,H_{0}^{2}\,\gamma_{10}
\;+\;\frac{3}{2}\,a^2\,
\frac{\bigl(H_{0}\,\gamma_{20}\bigr)^{2}}
     {2\,(1+\Omega_{0}) \;+\;3\,\gamma_{30}\;+\;\gamma_{40}}
\;,\\[6pt]
B\;=\;&
  \frac{\dot c\,a^2}{m_0^2}
  \;+\;4\,\mathcal{H}\,\frac{c\,a^2}{m_0^2}
  \;+\;8\,a^2\,\mathcal{H}\,H_{0}^{2}\,\gamma_{10}
  \;+\;a\,k^{2}\,
    \frac{\gamma_{40}+2\,\gamma_{50}}
         {2\,(1+\Omega_{0}) - 2\,\gamma_{40}}
    \,\bigl(H_{0}\,\gamma_{20}\bigr)
\notag\\[4pt]
&\quad
  -\;\frac{a\,k^2}
           {4\,(1+\Omega_{0}) + 6\,\gamma_{30} + 2\,\gamma_{40}}
  \Bigl[
    -3\,a^2\frac{\bigl(\rho_{Q}+P_{Q}\bigr)}{m_0^2}
    -3\,a\,\mathcal{H}\,H_{0}\,(4\,\gamma_{20})
    +(9\,\gamma_{30}+3\,\gamma_{40})\,(\dot{\mathcal{H}}-\mathcal{H}^{2})
\notag\\[4pt]
&\quad\quad
  +\;k^2\,(-3\,\gamma_{30}-\gamma_{40})
  +k^2\,(4\,\gamma_{50})
  \Bigr]
  \;+\;\frac{1}
           {1+\Omega_{0} + 2\,\gamma_{50}}
  \Bigl[
    2\,\mathcal{H}\,\gamma_{50}
    -(1+\Omega_{0})\,
      \frac{a\,H_{0}\,\gamma_{20}}
           {2\,(1+\Omega_{0}) + 3\,\gamma_{30} + \gamma_{40}}
  \Bigr]
\notag\\[4pt]
&\quad\quad\times
  \Bigl[
    -\,\frac{a^2c}{m_0^2}
    \;-\;2\,a^2\,H_{0}^{2}\,\gamma_{10}
    \;+\;\tfrac{3}{2}\,a\,\mathcal{H}\,H_{0}\,\gamma_{20}
  \Bigr],
\\[6pt]
C\;=\;&
  \mathcal{H}\,\frac{\dot c\,a^2}{m_0^2}
  \;+\;(6\,\mathcal{H}^{2}-2\,\dot{\mathcal{H}})\,\frac{c\,a^2}{m_0^2}
  \;+\;6\,\mathcal{H}^{2}\,H_{0}^{2}\,\gamma_{10}\,a^2
  \;+\;2\,\dot{\mathcal{H}}\,H_{0}^{2}\,\gamma_{10}\,a^2
  \;+\;\tfrac{3}{2}\,(\dot{\mathcal{H}}-\mathcal{H}^{2})^{2}\,(\gamma_{40}+3\,\gamma_{30})
\notag\\[-4pt]
&\quad
  +\;\frac{9}{2}\,\mathcal{H}\,H_{0}\,\gamma_{20}\,a\,(\dot{\mathcal{H}}-\mathcal{H}^{2})
  \;+\;\frac{a}{2}\,H_{0}\,\gamma_{20}\,\bigl(3\,\ddot{\mathcal{H}} -12\,\mathcal{H}\,\dot{\mathcal{H}}+6\,\mathcal{H}^{3}\bigr)
\notag\\[-4pt]
&\quad
  -\;\frac{a\,H_{0}\,\gamma_{20}}
           {4\,(1+\Omega_{0}) + 6\,\gamma_{30} + 2\,\gamma_{40}}
  \Bigl[
      -3\,a^2\,\frac{\dot P_{Q}}{m_0^2}
      -3\,\mathcal{H}\,a^2\frac{\bigl(\rho_{Q}+P_{Q}\bigr)}{m_0^2}
      +3\,(\ddot{\mathcal{H}}-2\,\mathcal{H}\,\dot{\mathcal{H}})\,(\gamma_{40}+3\,\gamma_{30})
\notag\\[-2pt]
&\qquad\qquad
      +6\,\mathcal{H}\,(\dot{\mathcal{H}}-\mathcal{H}^{2})\,(3\,\gamma_{30}+\gamma_{40})
      -3\,a\,H_{0}\,\gamma_{20}\,(3\,\mathcal{H}^{2}+\dot{\mathcal{H}})
  \Bigr]
\notag\\[-4pt]
&\quad
  +\;\frac{1}
           {1+\Omega_{0} + 2\,\gamma_{50}}
  \Bigl[
      2\,\mathcal{H}\,\gamma_{50}
      \;-\;(1+\Omega_{0})\,
        \frac{a\,H_{0}\,\gamma_{20}}
             {2\,(1+\Omega_{0}) + 3\,\gamma_{30} + \gamma_{40}}
  \Bigr]
\notag\\[-2pt]
&\qquad
  \Bigl[
      -\tfrac12\,\frac{a^2\,\dot\rho_{Q}}{m_0^2}
      -\,\mathcal{H}\,\frac{a^2\,c}{m_0^2}
      -2\,a^2\,H_{0}^{2}\,\gamma_{10}
      -\tfrac{3}{2}\,a\,H_{0}\,\gamma_{20}\,(\dot{\mathcal{H}}-2\,\mathcal{H}^{2})
      -3\,\mathcal{H}\,(\dot{\mathcal{H}}-\mathcal{H}^{2})\,\bigl(\tfrac32\,\gamma_{30}+\tfrac12\,\gamma_{40}\bigr)
  \Bigr],
\\[6pt]
D\;=\;&
\frac{c\,a^2}{m_0^2}
\;-\;\tfrac12\,a\,\mathcal{H}\,H_{0}\,\gamma_{20}
\;+\;(\mathcal{H}^{2}-\dot{\mathcal{H}})\,(3\,\gamma_{30}+\gamma_{40})
\;+\;2\,\dot{\mathcal{H}}\,\gamma_{50}
\notag\\[-4pt]
&\quad
+\;\frac{a\,H_{0}\,\gamma_{20}}
         {4\,(1+\Omega_{0}) + 6\,\gamma_{30} + 2\,\gamma_{40}}
\bigl[
  4\,\mathcal{H}\,\gamma_{50}
  \;-\;2\,\mathcal{H}\,(3\,\gamma_{30}+\gamma_{40})
\bigr]
\notag\\[-4pt]
&\quad
+\;\frac{1}
         {1+\Omega_{0} + 2\,\gamma_{50}}
\Bigl[
  2\,\mathcal{H}\,\gamma_{50}
  - (1+\Omega_{0})
    \frac{a\,H_{0}\,\gamma_{20}}
         {2\,(1+\Omega_{0}) + 3\,\gamma_{30} + \gamma_{40}}
\Bigr]
\Bigl[
  \tfrac12\,a\,H_{0}\,\gamma_{20}
  +\tfrac32\,\mathcal{H}\,\gamma_{30}
  +\tfrac12\,\mathcal{H}\,\gamma_{40}
  -2\,\mathcal{H}\,\gamma_{50}
\Bigr]
\notag\\[-4pt]
&\quad
+\;\frac{\gamma_{40}+2\,\gamma_{50}}
         {2\,(1+\Omega_{0}) - 2\,\gamma_{40}}
\Bigl[
  \frac{a^2(\rho_{Q}+P_{Q})}{m_0^2}
  -\gamma_{40}(\dot{\mathcal{H}}-\mathcal{H}^{2})
  +a\,\mathcal{H}\,H_{0}\,\gamma_{20}
  +3\,\gamma_{30}(\mathcal{H}^{2}-\dot{\mathcal{H}})
\Bigr]
\notag\\[-2pt]
&\quad
+\;k^2
\Bigl[
  \tfrac12\,\gamma_{30}
  +\tfrac12\,\gamma_{40}
  +\frac{\gamma_{40}+2\,\gamma_{50}}
         {2\,(1+\Omega_{0}) - 2\,\gamma_{40}}
    (\gamma_{30}+\gamma_{40})
\Bigr],
\\[6pt]
E\;=\;&
\frac{\dot h}{2}\,\Biggl\{
  \frac{c\,a^2}{m_0^2}
  -\tfrac12\,a\,\mathcal H\,H_{0}\,(2\,\gamma_{20})
  +\tfrac12\,\gamma_{30}\,\bigl(k^2-3\,\dot{\mathcal H}+3\,\mathcal H^2\bigr)
  +\tfrac12\,\gamma_{40}\,\bigl(k^2-\dot{\mathcal H}+\mathcal H^2\bigr)
\notag\\[-4pt]
&\quad
  -\;a\,\frac{H_{0}\,\gamma_{20}}
           {4\,(1+\Omega_{0}) \;+\;6\,\gamma_{30}\;+\;2\,\gamma_{40}}
    \Bigl[
      -4\,\mathcal H\,(1+\Omega_{0})
      \;-\;2\,\mathcal H\,(3\,\gamma_{30}+\gamma_{40})
    \Bigr]
\notag\\[-4pt]
&\quad
  +\;\frac{1}
           {1+\Omega_{0}+2\gamma_{50}}
    \Bigl[
      2\mathcal H\gamma_{50}
      \;-\;(1+\Omega_{0})\,
         \frac{H_{0}\,\gamma_{20}}
              {2\,(1+\Omega_{0}) \;+\;3\,\gamma_{30}\;+\;\gamma_{40}}
    \Bigr]
    \Bigl[
      \mathcal H\,\bigl(1+\Omega_{0}\bigr)
      +\tfrac12aH_{0}\gamma_{20}
      +\tfrac32\mathcal H\gamma_{30}
      +\tfrac12\mathcal H\gamma_{40}
    \Bigr]
\Biggr\}
\notag\\[6pt]
&\quad
+\;a\,
\frac{H_{0}\,\gamma_{20}}
     {4\,(1+\Omega_{0}) \;+\;6\,\gamma_{30}\;+\;2\,\gamma_{40}}
\;\bigl(\frac{a^2\,\delta P_{m,\nu}}{m_0^2}\bigr)
+\;(\gamma_{40}+2\,\gamma_{50})\,
\frac{k}
     {2\,(1+\Omega_{0}) \;-\;2\,\gamma_{40}}
\;\bigl(\frac{a^2(\rho_{m,\nu} + P_{m,\nu})\, v_{m,\nu}}{m_0^2}\bigr)
\notag\\[4pt]
&\quad
-\;
\frac{\,\mathcal H\,\gamma_{50}}
     {1+\Omega_{0} \;+\;2\,\gamma_{50}}
\;\bigl(\frac{a^2\,\delta \rho_{m,\nu}}{m_0^2}\bigr).
\end{align}
\end{subequations}
\end{widetext}

\bibliography{reference}

\end{document}